\documentclass[11pt,bm,aps,nofootinbib,amsfonts,amssymb,preprintnumbers]{revtex4}

 \usepackage{here}

\usepackage[dvipdfmx]{graphicx}
\usepackage{amssymb,amsfonts,amsmath,bm,color,multirow,cases,empheq,hyperref}
\usepackage{mathrsfs}
\newcommand{\red}[1]{{\color{red} #1}}

\usepackage{enumerate}
\usepackage{ulem}
\usepackage{afterpage}

\newcommand{\cL}{{\cal L}}
\newcommand{\cF}{{\cal F}}

\hypersetup{%
 setpagesize=false,
 bookmarksnumbered=true,%
 bookmarksopen=true,%
 colorlinks=true,%
 linkcolor=blue,%
 citecolor=red}

\begin{document}




\title{Causality of Photon Propagation under Dominant Energy Condition \\in Non-linear Electrodynamics}

\author{Shinya Tomizawa}
\email{tomizawa@toyota-ti.ac.jp}

\author{Ryotaku Suzuki}
\email{sryotaku@toyota-ti.ac.jp}

\affiliation{Mathematical Physics Laboratory, Toyota Technological Institute\\
Hisakata 2-12-1, Nagoya 468-8511, Japan}
\date{\today}

\preprint{TTI-MATHPHYS-23}




\begin{abstract} 
Recently, various types of regular black hole model are reintroduced as the solution of the Einstein equations coupled with nonlinear electrodynamics (NED). In NED, it is known that photons do not propagate along the null geodesics of the spacetime geometry, but of so-called effective geometry, which suggests the possibility of so-called ``faster/slower than light" photons.
We study the relation between the causality of photons and the dominant energy condition (DEC) in some static and spherically symmetric black hole spacetimes in NED.
We show that if photon trajectories with a nonzero angular momentum are timelike in the spacetime geometry, DEC is always satisfied in static and spherically symmetric spacetimes in any NED  that admits the Maxwell limit, and vice versa, at least, in the weak field limit. Thus, this implies that in such NED, the violation of DEC admits the existence of faster than light photons.
\end{abstract}

\date{\today}
\maketitle



\section{Introduction}
It is well-known that in nonlinear electrodynamics (NED), photons do not propagate along null geodesics of the spacetime geometry, but rather of another geometry which is called an effective geometry~\cite{Boillat:1970gw,Novello:1999pg,Novello:2000km}
. 
In such a theory, even if photons move in Minkowski spacetime, they move while feeling  a ``virtual" curved spacetime, whereas other massless particles move on the light-cone  in Minkowski spacetime.
For instance, this occurs in the Euler-Heisenberg effective theory derived by one-loop quantum correction in QED~\cite{Dittrich:1998fy,Brezin:1971nd,Drummond:1979pp,Latorre:1994cv,Shore:1995fz,Barton:1989dq,Scharnhorst:1990sr} and in the Born-Infeld elecrodynamics~\cite{Plebansky,Gibbons:2000xe}. 
 Therefore there is a possibility of faster-than-light photons and slower-than-light photons,  namely, the light-cone of the effective geometry lies outside the light-cone of the spacetime geometry, and vice versa.
Moreover, It is remarkable that in general, the velocities of photons in NED are doubled, which occurs in the former theory but does not in the latter~\cite{Gibbons:2000xe}.

\medskip
Regular black holes (RBH), which have no singularity inside/outside an event horizon, have been studied as one of the candidates of quantum black holes since Bardeen~\cite{Bardeen:1968} proposed the first model of asymptotically flat, static and spherically symmetric black holes with a regular center.
Subsequently, other RBHs with the same symmetry and asymptotic structure were proposed by many researchers.
Remarkably, some of these are exact solutions to the Einstein equation coupled with a physical source of a magnetic monopole in NED.
 Recently, Fan and Wang~\cite{Fan:2016hvf} found a wide class of asymptotically flat, static and spherically symmetric RBH solution in a certain NED, which can considered to be the generalization of the Bardeen BH~\cite{Bardeen:1968} and the Hayward BH~\cite{Hayward:2005gi}.  
 
 \medskip
 The propagation of photons has been studied for the Ay\'on-Beato-Garc\'ia spacetime in Ref.~\cite{Novello:2000km}, for the Bardeen spacetime in Ref.~\cite{Stuchlik:2019uvf} and for the Hayward spacetime in Ref.~\cite{Toshmatov:2019gxg}. 
 The photon orbits are also studied in  rotating versions of several RBHs of NED~\cite{Kumar:2020ltt}. 
Moreover, in Ref.~\cite{Isomura:2023oqf}, we discuss photons moving around regular black holes of Fan and Wang, and find an unstable circular orbit of photons inside the event horizon. 
The purpose of this paper is to study the relation between the energy conditions and the causality  of photon propagation around regular black holes in such NEDs, and see whether the light-cone of the effective geometry lies outside/inside the light-cone of the spacetime geometry under the dominant energy condition (DEC).
Furthermore, we show that in the Born-Infeld theory, where DEC is satisfied everywhere, photons with a nonzero angular momentum are always timelike for a purely magnetic/electric field. 
Moreover, generalizing these to arbitrary NED with the Maxwell limit, we show that under DEC, photon trajectories in static and spherically symmetric spacetimes cannot be spacelike, at least, in the weak field. 
In addition, we show that in any NED with the Maxwell limit if photon trajectories with a nonzero angular momentum are timelike in the spacetime geometry and null in the corresponding effective geometry, 
DEC is always satisfied in static and spherically symmetric spacetimes, which 
means that the violation of DEC leads to the existence of spacelike photons, i.e., faster than light photons.

 \medskip
 In the following section, we give the brief review on NED. 
 In section~\ref{sec:EC}, we discuss what conditions the Lagrangian density of NED should satisfy in order that four energy conditions, null energy condition (NEC), weak energy condition (WEC), DEC and strong energy condition (SEC) are satisfied in a purely magnetic and a purely electric cases. 
 In section~\ref{sec:EG},  we briefly review the known results on the effective geometry which photons in NED feel during propagating in the spacetime.
 In section~\ref{sec:EX}, 
 we first consider photons moving around the regular black hole of Fan and Wang, which can be regarded as a solution in a certain NED.
We also consider Einstein-Born-Infeld theory and discuss the relation between the energy conditions and the speed of photons.
In section~\ref{sec:GN}, we discuss the general cases with the Maxwell limit.
 In section~\ref{sec:summary},  we summarize our results and discuss possible generalization.



\section{Brief review}
Let us consider the Lagrangian density for Einstein gravity coupled with one parameter NED, which is given by
\begin{eqnarray}
L=R-{\cal L}(\cal F), \label{eq:action}
\end{eqnarray}
where ${\cal L}$ is an arbitrary function of ${\cal F}:=F_{\mu\nu}F^{\mu\nu}$ with the field strength of the vector field $A_\mu$, i.e.,  $F_{\mu\nu}=\partial_\mu A_\nu-\partial_\nu A_\mu$.
From the action (\ref{eq:action}), the Einstein equations and the field equations for NED can be written as, respectively, 
\begin{eqnarray}
&&G^\mu{}_\nu=2T^\mu{}_\nu,\label{eq:Einstein}\\
&&\nabla_\mu \left(\cL_\cF F^{\mu\nu} \right)=0, \label{eq:NED}
\end{eqnarray}
where the energy momentum tensor for NED is given by
\begin{align}
T^\mu{}_\nu =\cL_\cF F^{\mu\alpha}F_{\nu\alpha}-\frac{1}{4}\delta^\mu{}_\nu \cL .\label{eq:Tmunu}
\end{align}
The static and spherically symmetric solution with a purely magnetic field is written as
\begin{eqnarray}
ds^2 &=& -f(r) dt^2+f(r)^{-1}dr^2+r^2 d\theta^2+r^2\sin^2\theta d\phi^2,\\
 f(r)&=&1-\frac{2m(r)}{r},\\
 A_\mu dx^\mu &=& Q_m \cos\theta d\phi,\quad \cF = \frac{2Q_m^2}{r^4},
\end{eqnarray}
where the magnetic charge $Q_m$  is defined by
\begin{eqnarray}
Q_m:=\frac{1}{4\pi}\int_S F, 
\end{eqnarray}
and  Eq.(\ref{eq:Tmunu}) can be written as 
\begin{align}
(T^\mu{}_\nu) = \frac{1}{2}{\rm diag} \left(-\frac{1}{2}\cL ,-\frac{1}{2}\cL ,\cF \cL_\cF-\frac{1}{2}\cL ,\cF \cL_\cF-\frac{1}{2}\cL \right).\label{eq:emtensor}
\end{align}
From Eq.~(\ref{eq:Einstein}), the $(t,t)$, $(r,r)$ components and $(\theta,\theta)$, $(\phi,\phi)$ components are written as, respectively,  
\begin{eqnarray}
 &&\cL(\cF(r))= \frac{4m'}{r^2},\\
  &&\cL_\cF(\cF(r)) = -\frac{2r m''- 4 m'}{\cF^2 r^6} =  -\frac{r^2(r m''- 2 m')}{2Q_m^4} .
\end{eqnarray}

\medskip
Moreover, the static and spherically symmetric solution with a purely electric field is written as
\begin{eqnarray}
ds^2 &=& -f(r) dt^2+f(r)^{-1}dr^2+r^2 d\theta^2+r^2\sin^2\theta d\phi^2,\\
 f(r)&=&1-\frac{2m(r)}{r},\\
 A_\mu dx^\mu &=& a(r)dt.
\end{eqnarray}
The electric charge $Q_e$  is defined by
\begin{eqnarray}
Q_e:=\frac{1}{4\pi}\int_S \cL_\cF *F, 
\end{eqnarray}
and  Eq.(\ref{eq:Tmunu}) can be written as 
\begin{align}
(T^\mu{}_\nu) = \frac{1}{2}{\rm diag} \left(\cL_\cF\cF-\frac{1}{2}\cL ,\cL_\cF\cF-\frac{1}{2}\cL ,-\frac{1}{2}\cL ,-\frac{1}{2}\cL \right).\label{eq:emtensor}
\end{align}

\section{Energy conditions in NED}\label{sec:EC}
In this section we consider the conditions which the Lagrangian $\cL(\cF)$ should satisfy in order that four energy conditions, 
NEC, WEC, DEC and SEC are satisfied in a purely magnetic case ($\cF> 0$) and a purely electric case ($\cF< 0$).

\subsection{Purely magnetic case}
In this spacetime, the energy-momentum tensor~(\ref{eq:emtensor}) leads to the following energy conditions.

\begin{itemize}
\item NEC,  ``$T_{\mu\nu}k^\mu k^\nu \geq 0$ for any null vectors $k^\mu$", is equivalent with 
\begin{eqnarray}
 \cF \cL_\cF \geq 0. \label{eq:nec}
\end{eqnarray}

\item WEC, ``$T_{\mu\nu}v^\mu v^\nu \geq 0$ for any timelike vectors $v^\mu$",  is equivalent with
\begin{align}
 \cL \geq 0 \quad {\rm and} \quad \cF \cL_\cF \geq 0. \label{eq:wec}
\end{align}

\item DEC, ``$T_{\mu\nu}v^\mu v^\nu \geq 0$ and $J_\mu J^\mu \leq 0$ for any timelike vectors $v^\mu$ and the current $J^\mu:=-T^\mu{}_\nu v^\nu$",  is equivalent with
\begin{align}
\cL \geq 0,\quad  \cF \cL_\cF \geq  0 \quad {\rm and}\quad \cL - \cF \cL_\cF \geq 0. \label{eq:dec}
\end{align}

\item SEC, ``$\left(T_{\mu\nu}-\frac{1}{2}T^\lambda{}_\lambda g_{\mu\nu} \right) v^\mu v^\nu \geq 0$ for any timelike vectors $v^\mu$ is equivalent with
\begin{align}
\cL \geq 0, \quad \cF \cL_\cF \geq 0 \quad {\rm and} \quad 2\cF \cL_\cF -\cL \geq 0.
\end{align}
\end{itemize}

\subsection{Purely electric case}

\begin{itemize}
\item NEC,  ``$T_{\mu\nu}k^\mu k^\nu \geq 0$ for any null vectors $k^\mu$", is equivalent with 
\begin{eqnarray}
 \cF \cL_\cF \leq 0. \label{eq:nec2}
\end{eqnarray}

\item WEC, ``$T_{\mu\nu}v^\mu v^\nu \geq 0$ for any timelike vectors $v^\mu$",  is equivalent with
\begin{align}
 \cF \cL_\cF \leq 0 \quad {\rm and} \quad 2\cL_\cF \cF-\cL \leq 0. \label{eq:wec2}
\end{align}

\item DEC, ``$T_{\mu\nu}v^\mu v^\nu \geq 0$ and $J_\mu J^\mu \leq 0$ for any timelike vectors $v^\mu$ and the current $J^\mu:=-T^\mu{}_\nu v^\nu$",  is equivalent with
\begin{align}
 \cF \cL_\cF \leq 0, \quad {\rm and} \quad \cL_\cF \cF-\cL \leq 0. \label{eq:dec2}
\end{align}

\item SEC, ``$\left(T_{\mu\nu}-\frac{1}{2}T^\lambda{}_\lambda g_{\mu\nu} \right) v^\mu v^\nu \geq 0$ for any timelike vectors $v^\mu$,   is equivalent with
\begin{align}
 \cF \cL_\cF \leq 0 \quad {\rm and} \quad \cL \leq 0.
\end{align}
\end{itemize}

\section{Effective geometry and timelike photons} \label{sec:EG}

As mentioned previously, in NED given by the Lagrangian $\cL(\cF)$, photons do not propagate along null geodesics in the spacetime geometry, but rather in the corresponding effective geometry~\cite{Novello:1999pg}. 
In~Ref.\cite{Novello:1999pg}, by using the Hadamard method, the propagation of low-energy photons can be described by the evolution of the wave front, i.e., characteristic surface $S={\rm const}.$, across which the electromagnetic field is continuous but the first derivative is not. 
In Ref.~\cite{Stuchlik:2019uvf}, the alternative method, the eikonal approximation for photons, was used. 
Under the short-wave approximation in NED,  where the Faraday tensor can be regarded as local plane waves, 
\begin{eqnarray}
F_{\mu\nu}=\left( F_{\mu\nu}^{(0)}+\frac{\varepsilon}{i} F_{\mu\nu}^{(1)}+{\cal O}(\epsilon^2)+\cdots \right) e^{\frac{i}{\varepsilon}S} \quad (\varepsilon \ll 1)
\end{eqnarray}
from Eq.~(\ref{eq:NED}) and the Bianchi equations, one can show that the gradient of the phase $S$, $k_\mu:=\nabla_\mu S$,  must satisfy
\begin{align}
 \tilde{g}^{\mu\nu} k_\mu k_\nu =0,
\end{align}
where $\tilde{g}_{\mu\nu}$ is the metric of the effective geometry, which is given by
\begin{align}
\tilde{g}^{\mu\nu} = g^{\mu\nu} - \frac{4\cL_{\cF\cF}}{\cL_\cF} F^\mu{}_\alpha F^{\alpha \nu},\label{eq:eff-geom}
\end{align}
with  ${\cal L}_{\cal F F}:=d^2{\cal L}/d{\cal F}^2$.
For a static and spherically symmetric spacetime with a magnetic charge, the corresponding effective metric is denoted by
\begin{align}
(\tilde{g}_{\mu\nu}^{(m)}) = {\rm diag}\left(-f,\frac{1}{f},\frac{r^2}{\Phi^{(m)}},\frac{r^2\sin^2\theta}{\Phi^{(m)}}\right),
\end{align}
where $\Phi^{(m)}:= 1+2 \cL_{\cF\cF} \cF/\cL_\cF|_{\rm magnetic}$. 
On the other hand, in the electric case,  the effective metric is given by
\begin{align}
(\tilde{g}_{\mu\nu}^{(e)} )= {\rm diag}\left(-\frac{f}{\Phi^{(e)}},\frac{1}{f \Phi^{(e)}},r^2,r^2\sin^2\theta\right),
\end{align}
where $\Phi^{(e)}:=1+2 \cL_{\cF\cF} \cF/\cL_\cF|_{\rm electric}$. 
As shown in Ref.~\cite{Toshmatov:2021fgm},  from the duality in NED for the same metric,
\begin{align}
\cL_\cF^2 \cF|_{\rm electric} = -\cF|_{\rm magnetic},
\quad \cL_\cF|_{\rm magnetic} = (\cL_\cF)^{-1}|_{\rm electric},
\end{align}
one can show
\begin{align}
 \Phi^{(m)} = \frac{1}{\Phi^{(e)}},
\end{align}
which means that two effective geometries for the electric and magnetic solutions
are related by the conformal transformation. 
\begin{align}
 \tilde{g}_{\mu\nu}^{(e)} = \Phi^{(m)}\tilde{g}_{\mu\nu}^{(m)},
\end{align}
and hence have the same causal structure.
Thus, although the effective metrics for both the electrically and magnetically charged spacetimes are different, the photon trajectories coincide in both effective geometries. 
Therefore, it is sufficient to discuss the magnetic solution only. 
Let us denote $\Phi^{(m)}$ with $\Phi$ simply as
\begin{align}
\tilde{g}^{\mu\nu} = {\rm diag}\left(-\frac{1}{f},f,\frac{\Phi}{r^2},\frac{\Phi}{r^2\sin^2\theta}{\Phi}\right).
\end{align}
where
\begin{align}
 \Phi := 1 + \frac{2\cF \cL_{\cF\cF}}{\cL_\cF}.
\end{align}
If $k_\mu$ satisfies $\tilde{g}^{\mu\nu} k_\mu k_\nu=0$, then the norm of $k_\mu$ in the spacetime geometry is given by
\begin{align}
 g^{\mu\nu} k_\mu k_\nu = -\frac{2\cF \cL_{\cF\cF}}{r^2 \cL_\cF}  \left((k_\theta )^2+\frac{(k_\phi)^2}{\sin^2\theta}\right).
 \label{eq:kk}
\end{align}
Thus, under NEC and WEC, $\cF \cL_\cF\ge 0$, and if 
\begin{align}
  \cL_{\cF\cF} < 0,\label{eq:LFF}
\end{align}
 the trajectories of photons with $(k^\theta,k^\phi)\not=(0,0)$ can become timelike, namely,  photons propagate on the timelike  characteristic surfaces of $S=const$ because the gradient, $k^\mu=\nabla^\mu S$ is spacelike. 
We should note that for photons moving in the radial direction, the trajectories are also null even in the spacetime geometry.

\section{Examples} \label{sec:EX}
We consider two examples of BHs in NED, regular Fan-Wang (FW) black holes and Eisntein-Born-Infeld BHs.
\subsection{Regular black holes} \label{sec:FW}
As the first example, we consider the magnetic FW black holes~\cite{Fan:2016hvf},  which are solutions to Einstein equations coupled with NED given by the Lagrangian
\begin{align}
 {\cal L}({\cal F})=\frac{4\mu}{\alpha}\frac{(\alpha{\cal F})^{\frac{\nu+3}{4}}}{  \left(1+(\alpha{\cal F})^{\frac{\nu}{4}} \right)^{\frac{\mu+\nu}{\nu}}  },\label{eq:LF-FW-magnetic}
\end{align}
where we note $\cF>0$ for a purely magnetic case. 
The Bardeen BHs~\cite{Bardeen:1968} and Hayward BHs~\cite{Hayward:2005gi} are solutions in NED with $(\mu,\nu)=(3,2)$ and $(\mu,\nu)=(3,3)$, respectively.
In terms of $x:=(\alpha \cF)^{\nu/4}$, the energy conditions are denoted by
\begin{itemize}
\item {\rm NEC} $\Longleftrightarrow {\cal F}{\cal L}_{\cal F}>0 \Longleftrightarrow  \varepsilon_{\rm NEC}:=\nu+3 -(\mu-3)x >0$,
\item {\rm WEC} $\Longleftrightarrow  {\cal F}{\cal L}_{\cal F}>0$, ${\cal L}>0  \Longleftrightarrow     \varepsilon_{\rm NEC}>0$,
\item  {\rm DEC} $\Longleftrightarrow  {\cal F}{\cal L}_{\cal F}>0$, ${\cal L}>0$, ${\cal L}-{\cal F}{\cal L}_{\cal F}>0$  $\Longleftrightarrow  \varepsilon_{\rm NEC}>0,\  \varepsilon_{\rm DEC}:=1-\nu + (\mu+1) x>0$,
\item {\rm SEC} $\Longleftrightarrow   \cF \cL_\cF > 0, \ \cL > 0, \ 2\cF \cL_\cF -\cL > 0$ $\Longleftrightarrow  \varepsilon_{\rm NEC}>0,\ \varepsilon_{\rm SEC}:= 1+\nu -(\mu-1) x>0$,
\end{itemize}
where we note that the Lagrangian $\cL$ is always positive for magnetic black holes and $\varepsilon_{\rm NEC}$ is always positive for regular FW black holes with $\mu= 3,\ \nu \ge 1$, i.e., NEC and WEC are always satisfied.  
\medskip
The condition that  timelike photons exist under WEC is written as 
\begin{align}
&g^{\mu\nu}k_\mu k_\nu>0 \Longleftrightarrow \cL_{\cF\cF} <0  \nonumber\\
 &                        \hspace{2cm}            \Longleftrightarrow \varepsilon_{\gamma}:= (\mu+1)(\mu-3)x^2 -((3\nu+2)\mu+\nu^2-2\nu+6)x +(\nu+3)(\nu-1) <0.
\end{align}

In what follows, we classify four specific cases (i) $\mu=3,\nu=1$, (ii) $\mu>3,\nu=1$, (iii) $\mu=3,\nu>1$ and (iv) $\mu>3,\nu>1$.
\begin{itemize}
\item[(i)] $\mu=3,\nu=1$:
\begin{eqnarray}
\varepsilon_{\rm NEC}=4,\quad \varepsilon_{\rm DEC}=4x ,\quad \varepsilon_{\rm SEC}=2-2x. 
\end{eqnarray}
Thus, we can see from these that NEC, WEC, and DEC are satisfied everywhere but SEC are satisfied only for $x\ge 1$, i.e., not satisfied in the neighborhood of the regular center of black holes.  
On the other hand, from 
\begin{eqnarray}
\varepsilon_\gamma=-20x<0,
\end{eqnarray}
we find 
\begin{eqnarray}
g^{\mu\nu}k_\mu k_\nu>0, 
\end{eqnarray}
which means photons are timelike.

\item[(ii)] $\mu>3,\nu=1$:
\begin{eqnarray}
\varepsilon_{\rm NEC}=4-(\mu-3)x,\quad \varepsilon_{\rm DEC}=(\mu+1)x ,\quad \varepsilon_{\rm SEC}=2-(\mu-1)x.
\end{eqnarray}
From this, we can show that NEC, WEC, DEC, and SEC are all satisfied for $x<\frac{2}{\mu-1}$, NEC, WEC, DEC only are satisfied  $\frac{2}{\mu-1}<x<\frac{4}{\mu-3}$, and all energy conditions are not satisfied for $x>\frac{4}{\mu-3}$.
Since in this case, $\varepsilon_\gamma$ is written as 
\begin{eqnarray}
\varepsilon_\gamma=(\mu+1)[(\mu-3)x-5]x,
\end{eqnarray}
we can summarize the energy conditions and photon causality  in Table~\ref{tab:nu1}.

\begin{table}[H]
\centering
  \begin{tabular}{|c||c|c|c|c|}  \hline
    $x$  & $0<x<\frac{2}{\mu-1}$ & $\frac{2}{\mu-1}<x<\frac{4}{\mu-3}$ &  $\frac{4}{\mu-3}<x<\frac{5}{\mu-3}$  &$\frac{5}{\mu-3}<x<\infty $ \\ \hline\hline
          $\varepsilon_{\rm NEC} $          &  $+$                           &    $+$                                             &   $-$   & $-$  \\ \hline
            $\varepsilon_{\rm DEC} $          &  $+$                           &    $+$                                             &   $+$   & $+$  \\ \hline
            $\varepsilon_{\rm SEC} $          &  $+$                           &    $-$                                             &   $-$   & $-$  \\ \hline
              $\varepsilon_\gamma$ or $\cL_{\cF\cF} $          &  $-$                           &    $-$                                             &   $-$   & $+$  \\ \hline\hline
                     ${\rm NEC} $          &  Yes                           &    Yes                                             &   No   & No  \\ \hline
           ${\rm WEC} $          &  Yes                           &   Yes                                             &   No   & No  \\ \hline
            ${\rm DEC} $          &  Yes                           &   Yes                                             &   No   & No  \\ \hline
            ${\rm SEC} $          &  Yes                           &    No                                             &   No   & No  \\ \hline
           $g^{\mu\nu}k_\mu k_\nu $          &  $+$                           &    $+$                                             &   $-$   & $+$  \\ \hline
  \end{tabular}
    \caption{The energy conditions and photon causality for  $\mu>3,\nu=1$.}
    \label{tab:nu1}

\end{table}

\item[(iii)] $\mu=3,\nu>1$:
\begin{eqnarray}
\varepsilon_{\rm NEC}=\nu+3,\quad \varepsilon_{\rm DEC}=1-\nu+4x ,\quad \varepsilon_{\rm SEC}=1+\nu-2x.
\end{eqnarray}
From this, both NEC and WEC are satisfied everywhere. 
DEC and SEC are satisfied for $\frac{\nu-1}{4}<x<\frac{\nu+1}{2}$, DEC is satisfied but SEC is not for $x>\frac{\nu+1}{2}$, and SEC is satisfied but DEC is not for  $x<\frac{\nu-1}{4}$.
For  $\mu=3$, $\varepsilon_\gamma$ becomes 
\begin{eqnarray}
\varepsilon_\gamma=(\nu+3)[(\nu-1)-(\nu+4)x].
\end{eqnarray}
Therefore, we can summarize the energy conditions and photon causality  in Table~\ref{tab:mu3}.

\begin{table}[H]
\centering
  \begin{tabular}{|c||c|c|c|c|}  \hline
    $x$  & $0<x<\frac{\nu-1}{\nu+4}$ & $\frac{\nu-1}{\nu+4}<x<\frac{\nu-1}{4}$ &  $\frac{\nu-1}{4}<x<\frac{\nu+1}{2}$  &$\frac{\nu+1}{2}<x<\infty $ \\ \hline\hline
          $\varepsilon_{\rm NEC} $          &  $+$                           &    $+$                                             &   $+$   & $+$  \\ \hline
            $\varepsilon_{\rm DEC} $          &  $-$                           &    $-$                                             &   $+$   & $+$  \\ \hline
            $\varepsilon_{\rm SEC} $          &  $+$                           &    $+$                                             &   $+$   & $-$  \\ \hline
              $\varepsilon_\gamma$ or $\cL_{\cF\cF} $          &  $+$                           &    $-$                                             &   $-$   & $-$  \\ \hline\hline
                   ${\rm NEC} $          &  Yes                           &     Yes                                             &    Yes   & Yes  \\ \hline
               ${\rm WEC} $          &  Yes                           &    Yes                                             &   Yes   & Yes  \\ \hline
            ${\rm DEC} $          &  No                           &    No                                             &   Yes   & Yes  \\ \hline
            ${\rm SEC} $          &  Yes                           &    Yes                                             &   Yes   & No  \\ \hline
           $g^{\mu\nu}k_\mu k_\nu $          &  $-$                           &    $+$                                             &   $+$   & $+$  \\ \hline
  \end{tabular}
   \caption{The energy conditions and photon causality for  $\mu=3,\nu>1$.}
  \label{tab:mu3}
\end{table}

\item[(iv)] $\mu>3,\nu>1$:\\
NEC, WEC, DEC, and SEC are all satisfied  for $\frac{\nu-1}{\mu+1}<x<\frac{\nu+1}{\mu-1}$. 
NEC, WEC, DEC are satisfied but SEC is not for $\frac{\nu+1}{2}<x<\frac{\nu+3}{\mu-3}$, and NEC, WEC, SEC is satisfied but DEC is not for  $x<\frac{\nu-1}{\mu-3}$.
In general, $\varepsilon_\gamma=0$ can be solved as
\begin{eqnarray}
 x=x_\pm:=\frac{3 \mu  \nu +2 \mu +\nu ^2-2 \nu +6\pm \sqrt{(\mu+\nu) (5 \mu  \nu ^2+4 \mu  \nu +16 \mu +\nu ^3-4 \nu ^2+28 \nu)  }}{2(\mu+1)(\mu-3)}.
\end{eqnarray}
 From $\varepsilon_\gamma(\frac{\nu-1}{\mu+1})<0$,\ $\varepsilon_\gamma(\frac{\nu+3}{\mu-3})<0$,  we can see $0<x_-<\frac{\nu-1}{\mu+1}<\frac{\nu+3}{\mu-3}<x_+$. 
 Therefore we can summarize the energy conditions and photon causality in Table~\ref{tab:munu}.

\begin{table}[H]
\centering
  \begin{tabular}{|c||c|c|c|c|c|c|}  \hline
    $x$  & $0<x<x_-$ & $x_-<x<\frac{\nu-1}{\mu+1}$ &  $\frac{\nu-1}{\mu+1}<x<\frac{\nu+1}{\mu-1}$& $\frac{\nu+1}{\mu-1}<x<\frac{\nu+3}{\mu-3}$  &$\frac{\nu+3}{\mu-3}<x<x_+ $ & $x_+<x<\infty$\\ \hline\hline
          $\varepsilon_{\rm NEC} $          &  $+$                           &    $+$             & $+$                                &   $+$   & $-$ & $-$ \\ \hline
            $\varepsilon_{\rm DEC} $          &  $-$                           &    $-$              & $+$                                   &   $+$   & $+$  & $+$\\ \hline
            $\varepsilon_{\rm SEC} $          &  $+$                           &    $+$             & $+$                                    &   $-$   & $-$  & $-$\\ \hline
            $ \varepsilon_\gamma $ or $\cL_{\cF\cF} $          &  $+$                           &    $-$                          & $-$                       &   $-$   & $-$ & $+$ \\ \hline\hline
          ${\rm NEC} $          &  Yes                           &    Yes             & Yes                                &   Yes   & No & No \\ \hline
               ${\rm WEC} $          &  Yes                           &    Yes             & Yes                                &   Yes   & No & No \\ \hline
                       ${\rm DEC} $          &  No                           &     No              & Yes                                  &   Yes   & No  & No\\ \hline
            ${\rm SEC} $          &  Yes                           &    Yes             & Yes                                    &   No   & No  & No\\ \hline
           $g^{\mu\nu}k_\mu k_\nu $          &  $-$                           &    $+$               & $+$                                  &   $+$   & $-$  & $+$\\ \hline
  \end{tabular}
     \caption{The energy conditions and photon causality for  $\mu>3,\nu>1$.}
     \label{tab:munu}
\end{table}

\end{itemize}
In these cases 
we can conclude that if we assume that DEC is satisfied, photon trajectories can be timelike in the spacetime geometry, though  null in the effective geometry.

\subsection{Eisntein-Born-Infeld black holes}\label{sec:BI}
As the second example, we consider BHs in the Einstein-Born-Infeld theory~\cite{Born:1934gh}, whose Lagrangian density of NED is given by
\begin{eqnarray}
{\cal L}({\cal F})=-4\beta^2\left(1-\sqrt{1+\frac{{\cal F}}{2\beta^2}}\right),
\end{eqnarray}
which has the Maxwellian limit $\cL\simeq \cF$ in the weak field approximation $\cF\simeq 0$.
It is easy to show that the first and second derivatives have definite signatures
\begin{eqnarray}
{\cal L}_{\cal F}&=&\frac{1}{\sqrt{1+\frac{{\cal F}}{2\beta^2}}}>0,\\
{\cal L}_{\cal FF}&=&-\frac{1}{4\beta^2}\frac{1}{\sqrt{1+\frac{{\cal F}}{2\beta^2}}^3}<0.
\end{eqnarray}
It is sufficient to consider the purely magnetic case ${\cal F}>0$, in which all the energy conditions are satisfied  because it is obvious that 
\begin{itemize}
\item {\rm NEC} $\Longleftrightarrow {\cal F}{\cal L}_{\cal F} \ge 0$,
\item {\rm WEC} $\Longleftrightarrow  {\cal F}{\cal L}_{\cal F}\ge 0$, ${\cal L}\ge0$,
\item  {\rm DEC} $\Longleftrightarrow  {\cal F}{\cal L}_{\cal F}\ge 0$, ${\cal L}\ge 0$, ${\cal L}-{\cal F}{\cal L}_{\cal F} \ge0$,
\item  {\rm SEC} $\Longleftrightarrow  {\cal F}{\cal L}_{\cal F}\ge 0$, ${\cal L}\ge 0$, $2{\cal F}{\cal L}_{\cal F}-{\cal L}\ge 0$,
\end{itemize}
are satisfied everywhere. Therefore, from
\begin{eqnarray}
g^{\mu\nu}k_\mu k_\nu= -\frac{2{\cal F} {\cal L}_{\cal FF}}{r^2 {\cal L}_{\cal F}}  \left((k_\theta )^2+\frac{(k_\phi)^2}{\sin^2\theta}\right)\ge 0, 
\end{eqnarray}
in general, the photons propagating along null geodesics in the effective geometry  move along timelike curves  in the spacetime geometry except for photons moving in the radial direction with $(k^\theta,k^\phi)=(0,0)$, which propagate along null curves in both geometries.

\section{More general discussion}\label{sec:GN}
In more general cases, if we assume that the Lagrangian density with the Maxwell limit in the weak field limit has smoothness of $\cL$ at $\cF\to 0$, since 
\begin{eqnarray}
&&\cL \simeq   \cF+\frac{1}{2}\cL_{\cF\cF}(0)\cF^2+\cdots,\\
&&\cL_\cF\simeq 1 + \cL_{\cF\cF}(0) \cF +\cdots,
\end{eqnarray}
we can see 
\begin{eqnarray}
\cL-\cF \cL_\cF \simeq -\frac{1}{2}\cL_{\cF\cF}(0)\cF^2.
\end{eqnarray}
If we assume that DEC, $\cL\ge 0$, $\cF \cL\ge 0$, $\cL-\cF \cL_\cF \ge 0$, are satisfied in the weak field, the second-order derivative $\cL_{\cF\cF}(0) $ must be non-positive, which means that photons can be timelike or null  in the weak field such as at infinity.

\medskip
On the contrary, let us assume that photon propagation can be timelike, i.e.,
\begin{eqnarray*}
g^{\mu\nu}k_\mu k_\nu >0,
\end{eqnarray*}
which can be classified in two cases,  $\cL_{\cF\cF}<0$, $\cL_\cF>0$ and $\cL_{\cF\cF}>0$, $\cL_\cF<0$ from Eq.~(\ref{eq:kk}). The latter case does not admit the Maxwell limit, and hence we consider only the former case.
Moreover, from
\begin{eqnarray}
\partial_\cF(\cL-\cF \cL_\cF)=-\cF\cL_{\cF\cF}>0,
\end{eqnarray}
we find that 
\begin{eqnarray}
\cL-\cF \cL_\cF > (\cL-\cF \cL_\cF )|_{\cF=0}=0,\quad 
\cL > \cL|_{\cF=0}=0,
\end{eqnarray}
which means that DEC holds in such a region. 
In other words, the violation of DEC implies the existence of `faster than light' photons.

However, we should note that, without the Maxwell limit, the violation of DEC does not necessarily imply the existence of the spacelike propagation of photons as seen in examples shown in Table~ \ref{tab:mu3} and \ref{tab:munu}.

\section{Summary and Discussion}\label{sec:summary}
In this paper, we  have studied the causality  of photon propagation in NED when the energy conditions are satisfied.
As instances, we have considered the causality of photons around static and spherically symmetric Einstein-Born-Infeld BHs and well-known regular BHs such as Bardeen BHs, Hayward BHs, and Fan-Wang BHs, which can be regarded as static and spherically symmetric solutions to the Einstein equations coupled with NED. 
For such example, we have seen that as long as DEC is satisfied, the photon trajectories can be timelike in the spacetime geometry, though they are null in the effective geometry, i.e., the light-cone of the effective geometry does not lie outside the light-cone of the spacetime geometry. 

\medskip
In general, DEC can be interpreted as that the speed of energy flow of matter is always less than the speed of light. 
Hence, the existence of ``faster than light photons" in NED contradicts with DEC.
Indeed, we have shown that the violation of  DEC always leads to the existence of such photons, at least, in NED with the Maxwell limit, where we should note that this cannot necessarily be true in NED with no Maxwell limit  we have seen this in the examples of RBHs.

\medskip
In this paper, for simplicity, we have dealt with static and spherically symmetric spacetimes with a purely magnetic field or a purely electric field but we are not sure whether our results are also true for spacetimes with both fields, where an easy construction is shown in Ref.~\cite{Tsuda:2023tcs}, or rotating regular black holes~\cite{Torres:2022twv}. 
This deserves our future work. 

\medskip
Finally, we wish  to comment on the consistency with the results by Gibbons and Herdeiro in~\cite{Gibbons:2000xe}.
As discussed by them, the eigenvalues of the effective metric are proportional to
\begin{eqnarray}
\mu+\cF,\quad \mu+\cF, \quad \mu-\cF,\quad  \mu-\cF,
\end{eqnarray}
where $\mu$ is a root of the quadratic equation
\begin{eqnarray}
w\mu^2+\mu+\omega -w (\cF+{\cal G})=0,
\end{eqnarray}
where ${\cal G}:=F_{\mu\nu}*F^{\mu\nu}$.
Here, for the one-parameter Lagrangian density $\cL(F)$, the functions $w$ and $\omega$ are written as
\begin{eqnarray}
&&w:=\frac{\cL_{\cF\cF} \cL_{\cal G \cal G}-\cL_{\cF\cal G}^2}{\cL_\cF(\cL_{\cF\cF}+\cL_{\cal G\cal G})}=0,\\
&&\omega:=\frac{\cL_\cF +\cF (\cL_{\cF\cF}-\cL_{\cal G\cal G})+2 \cal G \cL_{\cF\cal G}}{\cL_{\cF\cF}+\cL_{\cal G\cal G}} =\cF +\frac{\cL_\cF}{\cL_{\cF\cF}},
\end{eqnarray}
therefore, the function $\mu$ turns out to be
\begin{eqnarray}
\mu=-\cF -\frac{\cL_\cF}{\cL_{\cF\cF}}.
\end{eqnarray}
The velocities of photons, the ratio of spacelike to timelike eigenvalues, are given by
\begin{eqnarray}
\left(1,\frac{\mu-x}{\mu+x},\frac{\mu-x}{\mu+x} \right)=\left(1,1+\frac{2\cF \cL_{\cF\cF}}{\cL_\cF} ,1+\frac{2\cF \cL_{\cF\cF}}{\cL_\cF}\right),
\end{eqnarray}
where the fact that the first component in the above equation is one means that there are two directions in which the light-cone in the effective geometry touches the usual light-cone in the spacetime geometry.
Under DEC, the condition of ${\cal L}_{\cal FF}<0$ is equivalent with that  the light-cone in the effective geometry does not lie outside  the light-cone in the spacetime geometry, i.e., photons are timelike or null in the spacetime geometry.

\acknowledgments
This work is supported by Toyota Technological Institute Fund for Research Promotion A. 
RS was supported by JSPS KAKENHI Grant Number JP18K13541. 
ST was supported by JSPS KAKENHI Grant Number 21K03560.




\end{document}